\documentclass[10pt,twocolumn,letterpaper]{article}
\usepackage{cvpr}
\usepackage{booktabs}
\usepackage{times}
\usepackage{epsfig}
\usepackage{graphicx}
\usepackage{amsmath}
\usepackage{amssymb}
\usepackage{xcolor}
\usepackage{multirow}
\usepackage{algorithm}
\usepackage{algpseudocode}
\usepackage{tabularx}
\usepackage{soul}
\usepackage{multirow}
\usepackage{adjustbox}
\usepackage{dblfloatfix}
\usepackage{caption} 
\usepackage{lipsum}
\usepackage{rotating}
\usepackage[belowskip=-1pt,aboveskip=0pt]{caption}
\usepackage[font=footnotesize,skip=0pt]{caption}
\usepackage{float}
\usepackage{subcaption}
\usepackage{footnote}
\usepackage{enumitem}

\usepackage[sort,square,numbers]{natbib}

\usepackage[breaklinks=true,bookmarks=false]{hyperref}

\cvprfinalcopy 

\def\cvprPaperID{****} 

\begin{document}

\title{Hierarchical Residual Attention Network for Single Image Super-Resolution}
\author{Parichehr Behjati\\
Computer vision center\\
Barcelona, Catalonia Spain \\
{\tt\small pbehjati@cvc.uab.es} 
\and
Pau Rodr\'{i}guez\\
Element AI\\
Montreal, Canada\\  
\and
Armin Mehri\\
Computer Vision Center\\
Barcelona, Catalonia Spain \\
\and
Isabelle Hupont\\
Herta Security\\
Barcelona, Catalonia Spain \\
\and
Carles Fernández Tena\\
Herta Security\\
Barcelona, Catalonia Spain \\
\and
Jordi Gonz\`alez\\
Computer Vision Center\\
Barcelona, Catalonia Spain \\
}


\maketitle


\newcommand{\architecture}{HRAN}
\newcommand{\resgroup}{RFG}
\newcommand{\RAFG}{RAFG}

\newcommand{\resblock}{residual block}
Convolutional neural networks are the most successful models in single image super-resolution. Deeper networks, residual connections, and attention mechanisms have further improved their performance. However, these strategies often improve the reconstruction performance at the expense of considerably increasing the computational cost. This paper introduces a new lightweight super-resolution model based on an efficient method for residual feature and attention aggregation. In order to make an efficient use of the residual features, these are hierarchically aggregated into feature banks for posterior usage at the network output. In parallel, a lightweight hierarchical attention mechanism extracts the most relevant features from the network into attention banks for improving the final output and preventing the information loss through the successive operations inside the network. Therefore, the processing is split into two independent paths of computation that can be simultaneously carried out, resulting in a highly efficient and effective model for reconstructing fine details on high-resolution images from their low-resolution counterparts. Our proposed architecture surpasses state-of-the-art performance in several datasets, while maintaining relatively low computation and memory footprint.

\section{Introduction}
\begin{figure}[!t]
\centering

\small
\begin{tabular}{ccc}
\includegraphics[width=.27\linewidth, height=4.5cm]{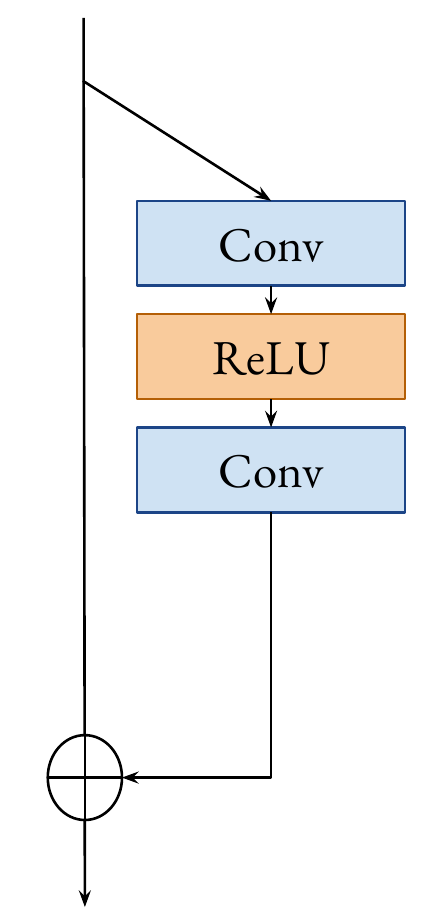} & \includegraphics[width=.28\linewidth, height=4.5cm]{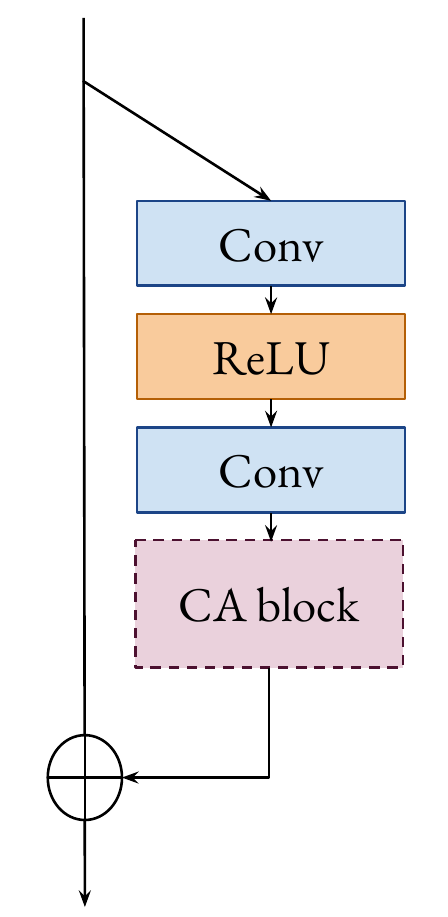}&  \includegraphics[width=.41\linewidth, height=4.5cm]{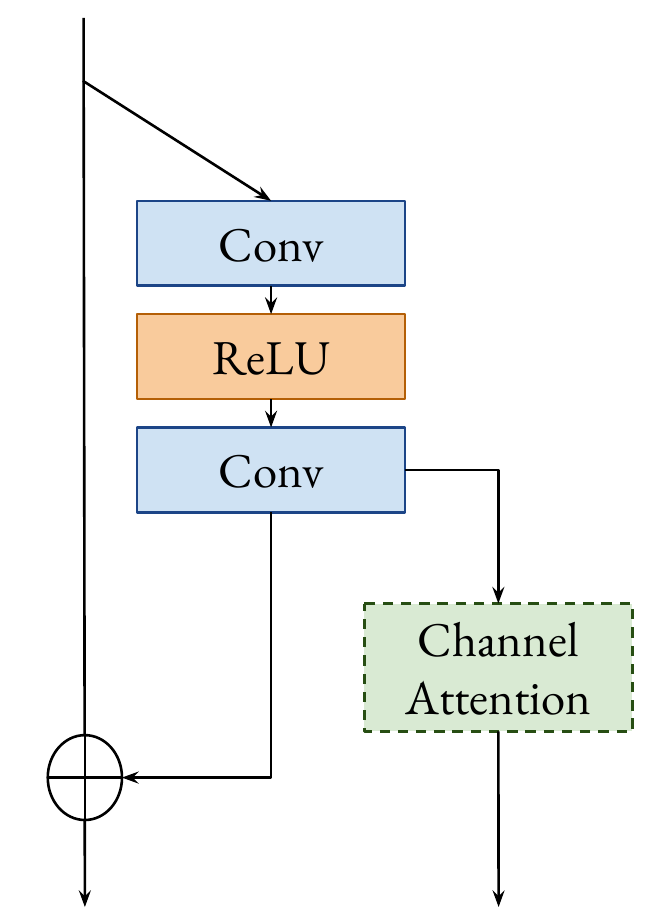}  \\
(a) & (b) & (c) 
\end{tabular}
\caption{\small (a)~Basic residual block without attention mechanisms. (b)~Channel-wise attention (CA) residual block proposed in previous works. (c) Our proposed channel attention mechanism with its own dedicated computational path.}
\label{tab:blocks}
\end{figure}

Single image super-resolution (SISR) is the task of reconstructing a high-resolution image (HR) from its low-resolution version (LR). This task is ill-posed, as multiple HR images can map to the same LR input. To tackle this inverse problem, numerous techniques have been proposed, including early interpolation~\cite{zhang2006edge}, reconstruction-based methods~\cite{zhang2012single} or, more recently, learning-based methods~\cite{ahn2018fast,dai2019second,niu2020single}. The development of convolutional neural networks (CNNs) followed by residual learning have paved the way for numerous advances in SISR. Most CNN-based SR methods employ a massive number of stacked residual blocks, as shown in Figure~\ref{tab:blocks}~(a), which allows deeper networks and a great increase in performance.

\setlength\tabcolsep{1.5pt}
\begin{figure*}[!t]
    \centering
    \includegraphics[width=\linewidth]{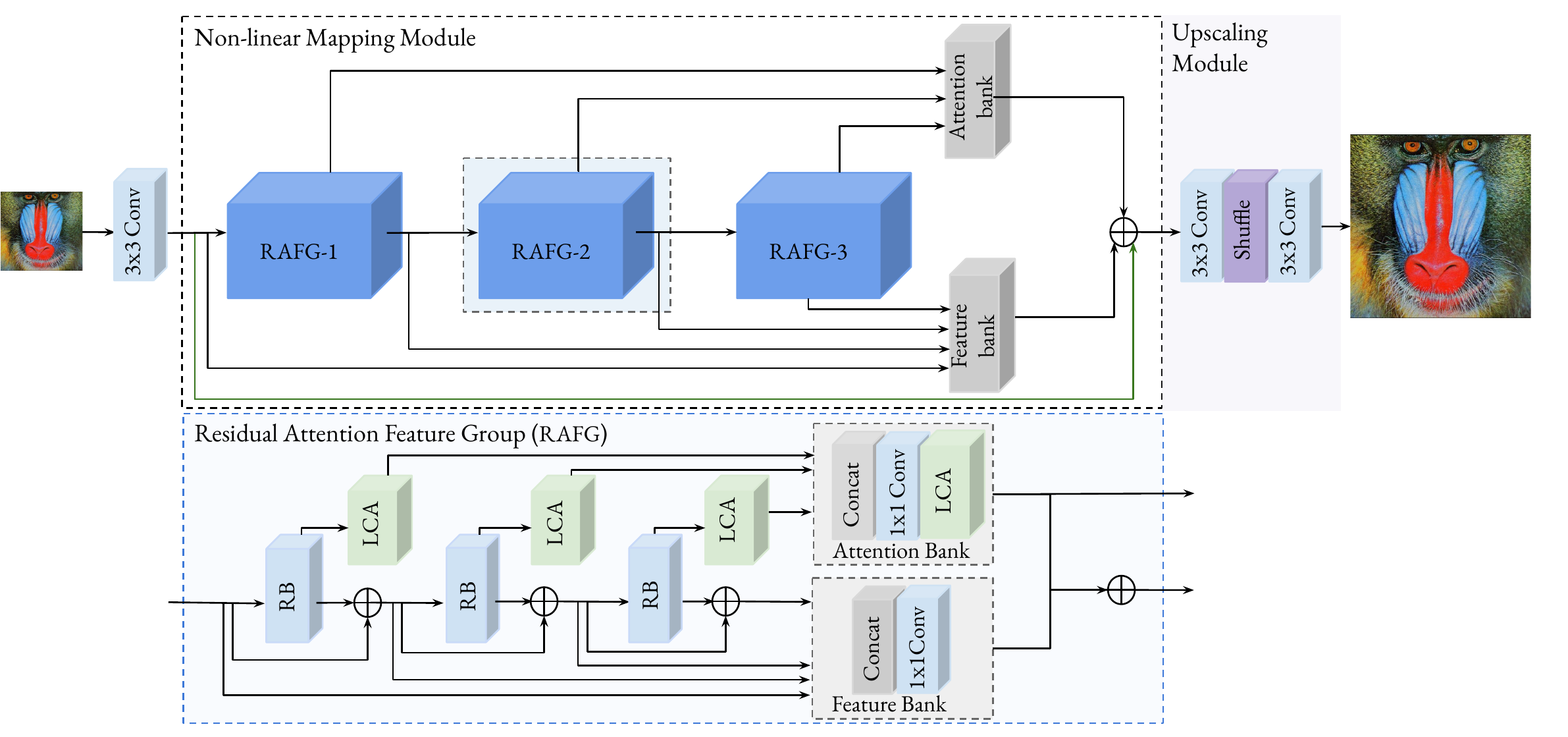}
    \captionsetup[figure]{font=small,skip=0pt}
    \caption{\small \textbf{Top}: Proposed Hierarchical Residual Attention Network (HRAN) architecture for SISR. \textbf{Bottom}: Residual Attention Feature Group (RAFG), containing residual blocks (RB) and lightweight channel attention (LCA) blocks}.
    \label{fig:main}
\end{figure*}
In spite of their outstanding performance, most SR deep networks still face some limitations. Usually, SR models are built by successively stacking residual blocks. Under this design, the residual features from preceding blocks must go through a long path to propagate until the final blocks, as these features are repeatedly merged with identity features to form more complex ones during transmission. Therefore, highly representative residual features are mostly computed locally, and lost in the residual addition during the learning process of the entire network. Moreover, very deep networks dramatically increase computational demands and memory consumption, which limits the use of modern architectures in real-world scenarios, such as mobile devices. To tackle these problems, 
some SR methods 
focus on lightweight architecture designs such as pyramid networks~\cite{lai2017deep}, recursive operations with weight sharing~\cite{ahn2018fast,choi2018lightweight,behjati2020overnet}, neural architecture search~\cite{hui2018fast,hui2019lightweight} and channel grouping~\cite{chu2019fast} to reduce the number of network parameters. 
Another problem of CNN-based SR methods as~\citep{lim2017enhanced,zhang2006edge} is that they are not explicitly designed 
to pay attention to 
the precious high-frequency information that must be reconstructed 
from the LR inputs. Previous works address this problem by re-weighting feature channels with channel attention~\cite{zhang2018image,niu2020single,muqeet2020multi}. However, channel attention may discard relevant details that will no longer be available at deeper levels of the neural architecture. Although attention-based approaches usually achieve better performance, they often increase model complexity and computational cost. Moreover, these methods were originally crafted for high-level vision problems such as image classification, so they do not take into account the particularities of SISR, like the need of preserving high-frequency components from LR inputs. In addition to this, how to effectively compute a multi-level feature representations for restoring high quality HR images within the network is of also crucial importance, yet it remains to be explored.

To confront these issues and address the problem of feature degradation, we introduce Residual Attention Feature Groups (RAFGs). The proposed \RAFG{} is composed of a collection of stacked residual blocks. The output of each residual block is linearly fused at the \RAFG{} output in order to take into account the whole feature hierarchy, to minimize the information loss during processing through the network, and to ease the gradient flow for optimization~\cite{huang2017densely}. This way, each residual block can focus on different aspects of the original LR input. 

However, using these stacked residual blocks solely is not sufficient to increase network's sensitivity towards information-rich channels. A number of previous approaches have tackled this challenge by performing in-place channel attention within the residual blocks, as in Figure~\ref{tab:blocks}~(b), to further boost the representational ability of the network. Nevertheless, such \textit{in-place} attention mechanisms may discard relevant details that will no longer be available at deeper levels of the architecture. To solve this problem, and in contrast to previous approaches, the proposed \RAFG{} is designed to combine lightweight channel attention blocks with residual blocks by keeping a dedicated computational path for attention modules. Figure~\ref{tab:blocks}~(c) shows that using \RAFG{}, attention modules are independent from the purely residual path, and parallel to it. As a result, they are able to attend to relevant features while preserving higher frequency details across the whole network.

To verify the effectiveness of the proposed \RAFG{}, we build upon them a deep architecture for SISR named Hierarchical Residual Attention Network (\architecture), illustrated in Figure~\ref{fig:main}. Our experiments show that our network outperforms previous state-of-the-art approaches in standard benchmarks, while maintaining low computation and memory requirements. In summary, the main contributions of this paper are as follows:

\begin{enumerate}
    \item We introduce a novel procedure for using channel attention mechanisms together with residual blocks, following two independent but parallel computational paths. The idea is to hierarchically aggregate their respective contributions across the network, using Residual Attention Feature Groups to facilitate the preservation of finer details.
    \item To demonstrate the potential of the proposed attention mechanism, we present the Hierarchical Residual Attention Network (HRAN), a deep  architecture for SISR grounded on the use of RAFGs. Extensive experiments on a variety of public datasets demonstrate the superiority of the proposed architecture over state-of-the-art models.
    \item As a consequence of the previous, we obtain a lightweight and efficient SISR model, in which features and attention are processed simultaneously, thus improving the performance while reducing memory and processing costs. 
\end{enumerate}

\section{Related Work}

\subsection{Evolution of Architectures for SISR}

\citet{dong2014learning} first proposed SRCNN, a shallow neural network to predict the non-linear relationship between bicubic upsamplings and HR images. SRCNN has a large number of operations compared to its depth, as it extracts features from upsampled LR images. In contrast, FSCRNN~\cite{dong2016accelerating} and ESPCN~\cite{lim2017enhanced} started to directly adopt the LR image as input by adding a single transposed convolution layer to produce the final HR output, which lead to a reduction in the number of operations when compared to SRCNN. These early methods are considered shallow models. 

Subsequent approaches have shown better performances than shallow models by using deep neural networks~\citep{urban2016deep}. VDSR~\cite{kim2016accurate} and IRCNN~\cite{zhang2017learning} improved the performance by increasing the network depth, using stacked convolutions with residual connections. \citet{lim2017enhanced} further expanded the network size and improved residual blocks by removing batch normalization layers. Furthermore, to effectively send information across layers, dense blocks~\cite{huang2017densely} were employed in several SR deep networks~\citep{zhang2018residual,tong2017image,wang2018fully}. In MemNet, \citet{tai2017memnet} designed memory blocks also based on dense connections. Meanwhile, to simplify the challenge of directly super-resolving the details, \citet{he2018cascaded} adopted a progressive structure to reconstruct HR images in a stage-by-stage upscaling manner. \citet{li2019feedback} and \citet{haris2018deep} incorporated the feedback mechanism into network designs for exploiting both LR and HR signals jointly. Recently, \citet{luolatticenet} proposed  butterfly structures in LatticeNet, combining residual blocks to further improve results. Nevertheless, despite their remarkable performance, such methods demand very high computational cost and memory requirements.

Numerous lightweight models have been proposed to alleviate the computational burden. For example, DRCN~\cite{kim2016deeply} was the first to apply recursive algorithms to SISR to reduce the number of parameters. \citet{kim2016deeply} revisited DRCN by combining recursive structures and residual blocks so as to improve performance with even fewer parameters. Likewise, \citet{ahn2018fast,behjati2020overnet} also joined residual connections and recursive layers to reduce the computational cost. \citet{chu2019fast} introduced Neural Architecture Search (NAS) strategies to automatically build an SR model given certain constraints.  \citet{zhu2019efficient} proposed CBPN, an efficient version of the DBPN network~\cite{haris2018deep}, which emphasizes high-resolution features present in LR images. Very recently, \citet{muqeet2020multi} proposed an architecture based on sequentially stacked residual blocks. All the aforementioned works demonstrate that lightweight SR networks are capable of providing good trade-offs between performance and number of parameters.

\subsection{Attention Mechanisms in SISR}

The aim of introducing attention mechanisms to neural networks is to re-calibrate the feature responses towards the most informative and important components of the inputs. In short, they help the network ignore irrelevant information and focus on the important one \cite{hu2018squeeze,rodriguez2018attend}.
For example, \citet{woo2018cbam} stacked spatial and channel-wise attention modules at multiple layers for image classification. 
Attention mechanisms have also been successfully applied to deep CNN-based image enhancement methods and, more particularly, to SISR. \citet{zhang2018image} first proposed a very deep residual channel attention network (RCAN), which integrated channel-wise attention into residual blocks and pushed the state-of-the-art performance of SISR. 

Only very recently, SISR works have started to investigate more sophisticated attention mechanisms, leaving room for further research in field. \citet{liu2020residual} proposed enhanced spatial attention (ESA), a combined solution for channel and spatial attention, which reduces the number of channels using 1$\times$1 convolutions and shrinks spatial dimensions via strided convolutions. Inspired by ESA, \citet{muqeet2020multi} presented a cost-efficient attention mechanism (CEA) introducing dilated convolutions with different filter sizes. \citet{niu2020single} proposed a layer and channel-wise attention mechanism to selectively capture more informative features. Meanwhile, \citet{mehri2020mprnet} proposed to combine both channel-wise and spatial attentions to learn more expressive context information. Another example is the work by \citet{zhao2020efficient}, which introduces a pixel-wise channel attention to produce 3D attention maps. 


\section{Hierarchical Residual Attention Network}
\label{sec:proposed_method}

In this section, we first provide an overview of the proposed Hierarchical Residual Attention Network (\architecture{}) for SISR. Then, we detail the components of a residual attention feature group (RAFG).

\subsection{HRAN Overview}
As shown in Figure~\ref{fig:main}, \architecture{} consists of a non-linear mapping module and a final upscaling and image reconstruction module.
Let's denote  $I_{LR}$ and $I_{SR}$ the input and output of \architecture, respectively. As recommended in \cite{ledig2017photo,lim2017enhanced}, we apply only one $3\times3$ convolutional layer to extract the initial features $F_0$ from the LR input image:

\begin{equation}
    F_{0}= \operatorname{Conv}_{3\times3}(I_{LR})
\end{equation}

Next, the extracted features $F_0$ are sent to the non-linear mapping module, which computes useful representations of the LR patch in order to infer its HR version. 
Note that we also introduce a long-range skip connection to grant access to the original information and facilitate the back-propagation of the gradients (green connection in Figure~\ref{fig:main}).

Let F be the output of the non-linear mapping module, further detailed in Section~\ref{sec_RAFGs}, which contains high resolution features. Then, an upscaling process grounded on a sub-pixel shuffle convolutional layer~\cite{shi2016real} is applied:

\begin{equation}
    F^{\uparrow} = \operatorname{sub-pixel shuffle}(F).
\end{equation}

The upscaled features $F^{\uparrow}$ are finally mapped into the SR image $I_{SR}$ via one convolutional layer:

\begin{equation}
    I_{SR} = \operatorname{Conv_{3\times3}}(F^{\uparrow}).
\end{equation}

To optimize \architecture{} and following previous works, we adopt the $L_1$ loss as a cost function for training. Given a training set with $N$ pairs of LR images and HR counterparts, denoted by \{${I^{i}_{LR}, I^{i}_{HR}}$\}$^N_{i=1}$,  \architecture{} is optimized to minimize the $L_1$ loss function:

\begin{equation}
 L(\boldsymbol{\theta}) = \frac{1}{N} \sum_{i=1}^{N}\left \| I_{LR} - I_{HR}\right\|_1,
\end{equation}

\noindent where $\boldsymbol{\theta}$ denotes the parameter set of \architecture. 

\subsection{Residual Attention Feature Group}
\label{sec_RAFGs}

The residual attention feature group (RAFG) is the core of the non-linear mapping module. It is designed to attend and preserve higher frequency details across the whole network. As shown in Figure~\ref{fig:main} (bottom), it is composed of two dedicated computational paths: (i) feature banks and (ii) attention banks. We detail each of these below.
\paragraph{Feature banks.} In essence, residual features are linearly combined at banks which are built by aggregating all the features from previous residual blocks. As a result, these banks capture detailed information from features across the whole architecture, thus reducing feature degradation and boosting the network's overall representational ability. Given that the systematic concatenation of features results in redundant information, we also incorporate 1$\times$1 bottleneck convolutions after feature aggregation.

Compared to a simple stacking of multiple residual blocks, the use of hierarchical banks enables a non-local exploitation of residual features. Although stacked residual blocks facilitate the task of training deeper SISR networks, residual feature information gradually disappears during the transmission and residual blocks play a very local role in the learning process of the whole network. This is because residual features from preceding blocks must go through a long path to propagate to subsequent ones.

In contrast, with feature banks, information from preceding residual blocks can be hierarchically propagated bottom-up without any loss or interference, thus leading to a more discriminative feature representation.


\begin{figure}[!t]
    \centering
    \includegraphics[width=0.7\linewidth]{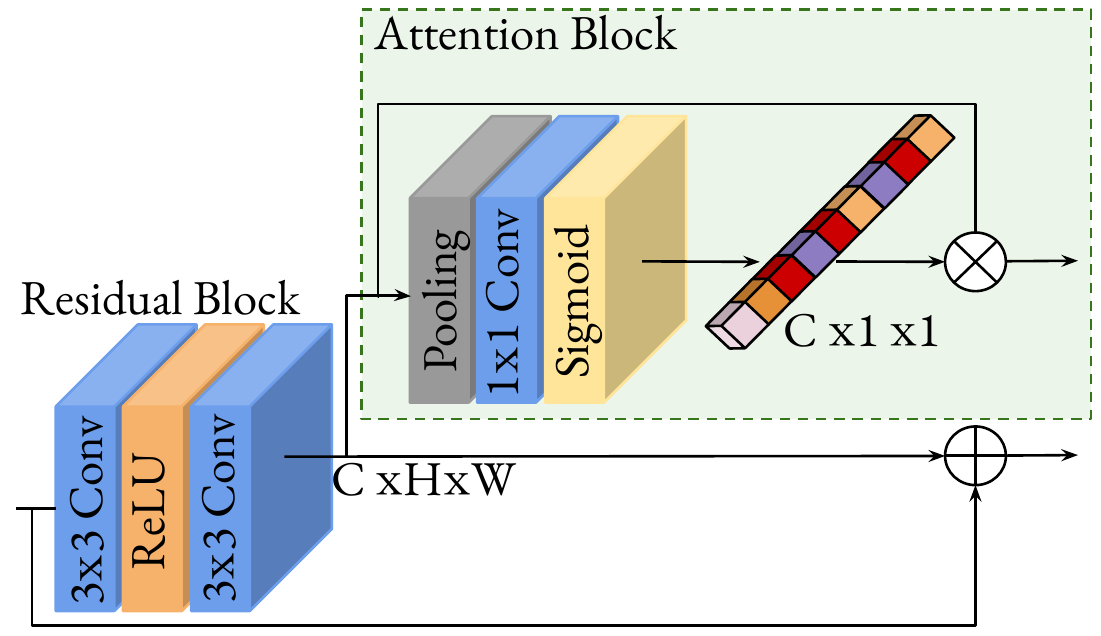}
    \caption{Proposed mechanism for extracting channel-wise attention from residual blocks.}
    \label{fig:attention}
\end{figure}
\paragraph{Lightweight Channel Attention (LCA) blocks.} To further improve the performance of \architecture, we also propose the use of LCA blocks.
Their structure is illustrated in Figure~\ref{fig:attention}. The attention recalibrates feature responses by implicitly modeling interdependencies between channels. In order to make it efficient, we propose to use a single 1x1 convolutional layer instead of the two 1x1 bottleneck convolutions commonly used in the literature~\cite{zhang2018image}. Formally, we denote $u = [u_1, ..., u_C]$ the input of the LCA block, 
which consists of $C$ feature maps $u_i$ of spatial size $H\times W$. These feature maps are first reduced in the spatial dimension by an average pooling operation:

\begin{equation}
    z_c = \frac{1}{H\times W}\sum_{i=1}^{H}\sum_{j=1}^{W}u_c(i,j),
\end{equation}

\noindent where $u_c(i,j)$ is the value at position $(i,j)$ of the $c$-th channel, and $z_c$ is the value obtained for channel $c$ 
after average pooling. To attend to the most relevant feature maps, we employ a gating mechanism with sigmoid activation:

\begin{equation}
    \alpha_c =\sigma (W_{LCA} * z_c),
\end{equation}

\noindent where $\sigma$ represents the sigmoid function, $z = [z_1, \ldots, z_C]$,
and $*$ denotes a linear transformation with $C\times C$ weights $W_{LCA}$. The final channel importance weights $\alpha=[\alpha_1, ..., \alpha_C]$ are used to re-scale the input $u$:
\begin{equation}
    \tilde{u}_c = \alpha_c . u_c
\end{equation}
\paragraph{Integrating LCA in \architecture{}.} The features extracted by a deep network contain different types of information in each channel. If we are able to increase the network's sensitivity to information-rich channels and make it focus on them, the efficiency of the network will be enhanced and its performance improved. In the literature, channel attention mechanisms have been one of the most successful strategies for this purpose. Previous approaches performed channel attention \textit{in-place} inside the residual blocks to further boost the representational ability of the network~\cite{hu2018squeeze}. This usually implied an element-wise product between the attention output and the residual block output. However, this in-place channel attention may discard relevant details which will no longer be available in deeper levels of the architecture.

In our approach, we propose to keep a dedicated computational path for attention modules, independent from the aggregation of residual features, and parallel to it. Figure \ref{fig:main}~(bottom) shows that a RAFG contains three residual blocks, the output of which are respectively sent to an LCA block before element-wise addition. 
Attention outputs are then globally aggregated as a bank of attentions, and the same happens with the residual features, which are globally combined as a bank of features. The attention bank additionally incorporates an extra LCA block, which helps to efficiently extract the most important components of the feature hierarchy. Finally, the outputs of both banks are combined together and sent to the subsequent RAFG. Likewise, the outputs of all the feature banks and attention banks are combined by a final, higher-level feature bank and attention bank at the output of the network.

\section{Experimental Results}
In this section, we first compare our \architecture{} with state-of-the-art algorithms on five benchmark datasets. We then analyze the contributions of the proposed channel attention mechanism and residual attention feature group.
\subsection{Settings}
\noindent \textbf{Datasets}. Following \cite{niu2020single,liu2020residual,muqeet2020multi}, we use 800 high-quality training images from DIV2K dataset as our training set. Several benchmark datasets are used for testing, namely Set5~\cite{bevilacqua2012low}, Set14~\cite{zeyde2010single}, B100~\cite{arbelaez2010contour}, and Urban100~\cite{arbelaez2010contour}, each with diverse characteristics. We perform our experiments with bicubic (BI) and blur-downscale (BD) degradation models. All the results are evaluated with two commonly used metrics: PSNR (peak-to-noise-ratio) and SSIM (structural similarity index), on the Y channel of the YCbCr space.\\ 
\noindent \textbf{Implementation details}. During training, data augmentation is carried out by means of random horizontal flips and $90^{\circ}$ rotation. At each training mini-batch, 64 LR RGB patches with size $64\times64$ are provided as inputs. We train an \architecture{} consisting of 3 RAFGs, each with 3 residual blocks (RBs) and 3 attention modules (LCAs). We train our models using the ADAM optimizer with learning rate set to the maximum convergent value (${10^{-3}}$), applying weight normalization in all convolutional layers. The learning rate is decreased by half every $2\times{10^5}$ back-propagation iterations. The proposed architecture has been implemented using PyTorch~\cite{paszke2017automatic}, and trained on NVIDIA 1080 Ti GPUs\footnote{The code is publicly available at: (blind for review process)}.

\begin{table*}[!t]

\caption{\small Average PSNR/SSIM values for models with the same order of magnitude of parameters.  Performance is shown for scale factors $\times$2, $\times$3 and $\times$4 with \textbf{BI} degradation. The number of parameters and of each method are indicated  under their name. The best performance is shown \textbf{highlighted} and the second best \underline{underlined}. }
\begin{adjustbox}{width=1\textwidth}
\setlength\arrayrulewidth{1.3pt}
\begin{tabular}{c@{~~}c@{~}c@{~~~~}c@{~~~~}c@{~~~~}c@{~~~~}c@{~~~}c@{~~}c@{~~~}c@{~~~}c@{~~~}c@{~~~}c@{~~~}c@{~~~}}
\toprule
{\textbf{Dataset}} &  {\begin{tabular}[c]{@{}c@{}}\textbf{Scale}\\ \end{tabular}}
 & {\begin{tabular}[c]{@{}c@{}}\textbf{VDSR}\\ \textbf{0.7M}\end{tabular}} & {\begin{tabular}[c]{@{}c@{}}\textbf{DRNN}\\ \textbf{0.3M}\end{tabular}} & {\begin{tabular}[c]{@{}c@{}}\textbf{MemNet}\\ \textbf{0.7M}\end{tabular}} & {\begin{tabular}[c]{@{}c@{}}\textbf{CARN}\\ \textbf{1.6M}\end{tabular}} & {\begin{tabular}[c]{@{}c@{}}\textbf{SRFBN\_s}\\ \textbf{0.5M}\end{tabular}} & {\begin{tabular}[c]{@{}c@{}}\textbf{IMDN}\\ \textbf{0.7}\end{tabular}} &
{\begin{tabular}[c]{@{}c@{}}\textbf{PAN}\\ \textbf{0.3M}\end{tabular}} &
{\begin{tabular}[c]{@{}c@{}}\textbf{MAFFSRN\_L}\\ \textbf{0.8M}\end{tabular}} &
  {\begin{tabular}[c]{@{}c@{}}\textbf{LatticeNet}\\ \textbf{0.8M}\end{tabular}} &
 {\begin{tabular}[c]{@{}c@{}}\textbf{MPRNet}\\ \textbf{0.5M}\end{tabular}} & 
 {\begin{tabular}[c]{@{}c@{}}\textbf{OverNet}\\ \textbf{0.9M}\end{tabular}} &
  {\begin{tabular}[c]{@{}c@{}}\textbf{Ours}\\ \textbf{0.9M}\end{tabular}}

  \\

\midrule
{\textbf{Set5}} & {\begin{tabular}[c]{@{}l@{}}$\times$2\\ $\times$3\\ $\times$4\end{tabular}} &  {\begin{tabular}[c]{@{}c@{}}37.53/0.9587\\ 33.66/0.9213\\ 31.35/0.8838\end{tabular}} & {\begin{tabular}[c]{@{}c@{}}37.74/0.9591\\ 34.03/0.9244\\ 31.68/0.8888\end{tabular}} & {\begin{tabular}[c]{@{}c@{}}37.78/0.9597\\34.09/0.9248\\31.74/0.8893\end{tabular}} & {\begin{tabular}[c]{@{}c@{}}37.76/0.9590\\ 34.29/0.9255\\ 32.13/0.8937\end{tabular}} & {\begin{tabular}[c]{@{}c@{}}38.02/0.9605\\ 34.20/0.9255\\ 31.98/0.8923\end{tabular}} & 
{\begin{tabular}[c]{@{}c@{}}38.00/0.9605\\34.36/0.9270\\32.21/0.8948\end{tabular}} &{\begin{tabular}[c]{@{}c@{}}38.00/0.9605\\34.40/0.9271\\32.13/0.8948\end{tabular}}  &
{\begin{tabular}[c]{@{}c@{}}38.07/0.9607\\34.45/0.9277\\32.20/0.8953\end{tabular}} &

{\begin{tabular}[c]{@{}c@{}}\underline{38.15/0.9610}\\34.53/0.9281\\32.30/0.8962\end{tabular}} &

{\begin{tabular}[c]{@{}c@{}}38.08/0.9608\\\underline{34.57/0.9285}\\ \underline{32.38/0.8969}\end{tabular}} &

{\begin{tabular}[c]{@{}c@{}}38.11/0.9610\\ 34.49/0.9267\\ 32.32/0.8956\end{tabular}}

& {\begin{tabular}[c]{@{}c@{}}\textbf{38.23/0.9616}\\ \textbf{34.62/0.9288} \\ \textbf{32.50/0.8976}\end{tabular}} \\ 
\midrule
{\textbf{Set14}} & {\begin{tabular}[c]{@{}l@{}}$\times$2\\ $\times$3\\ $\times$4\end{tabular}} &  {\begin{tabular}[c]{@{}c@{}}33.03/0.9124\\29.77/0.8314\\28.01/0.7674\end{tabular}} & {\begin{tabular}[c]{@{}c@{}}33.23/0.9136\\29.96/0.8349\\28.21/0.7720\end{tabular}} & {\begin{tabular}[c]{@{}c@{}}33.28/0.9142\\30.00/0.8350\\28.26/0.7723\end{tabular}} & {\begin{tabular}[c]{@{}c@{}}33.52/0.9166\\30.29/0.8407\\28.60/0.7806\end{tabular}} & {\begin{tabular}[c]{@{}c@{}}33.35/0.9156\\30.10/0.8372\\ 28.45/0.7779\end{tabular}} & 
{\begin{tabular}[c]{@{}c@{}}33.63/0.9177\\30.32/0.8417\\28.58/0.7811\end{tabular}} &{\begin{tabular}[c]{@{}c@{}}33.59/0.9181\\30.36/0.8423\\28.61/0.7822\end{tabular}}  &
{\begin{tabular}[c]{@{}c@{}}33.59/0.9177\\30.40/0.8432\\28.62/0.7822\end{tabular}} &

{\begin{tabular}[c]{@{}c@{}}33.78/0.9193\\30.39/0.8424\\28.68/0.7830\end{tabular}} &

{\begin{tabular}[c]{@{}c@{}}\underline{33.79/0.9196}\\30.42/\underline{0.8441}\\ 28.69/\underline{0.7841}\end{tabular}} &

{\begin{tabular}[c]{@{}c@{}}33.71/0.9179\\ \underline{30.47}/0.8436\\\underline{28.71}/0.7826\end{tabular}}

& {\begin{tabular}[c]{@{}c@{}}\textbf{33.81/0.9198}\\ \textbf{30.55/0.8449} \\ \textbf{28.76/0.7848}\end{tabular}} \\ 
 \midrule
{\textbf{B100}} & {\begin{tabular}[c]{@{}l@{}}$\times$2\\ $\times$3\\ $\times$4\end{tabular}} &  {\begin{tabular}[c]{@{}c@{}}31.90/0.8960\\28.82/0.7976\\27.29/0.7251\end{tabular}} & {\begin{tabular}[c]{@{}c@{}}32.05/0.8973\\28.95/0.8004\\27.38/0.7284\end{tabular}} & {\begin{tabular}[c]{@{}c@{}}32.08/0.8978\\28.96/0.8001\\27.40/0.7281\end{tabular}} & {\begin{tabular}[c]{@{}c@{}}32.09/0.8978\\29.06/0.8034\\27.58/0.7349\end{tabular}} & {\begin{tabular}[c]{@{}c@{}}32.20/0.9000\\29.11/0.8085\\ 27.60/0.7369\end{tabular}} & 
{\begin{tabular}[c]{@{}c@{}}32.19/0.8996\\29.09/0.8046\\27.56/0.7353\end{tabular}} &{\begin{tabular}[c]{@{}c@{}}32.18/0.8997\\29.11/0.8050\\27.59/0.7363\end{tabular}}  &
{\begin{tabular}[c]{@{}c@{}}32.23/0.9005\\29.13/0.8061\\27.59/0.7370\end{tabular}} &

{\begin{tabular}[c]{@{}c@{}}\underline{32.25/0.9005}\\29.15/0.8059\\27.62/0.7367\end{tabular}} &

{\begin{tabular}[c]{@{}c@{}}\underline{32.25}/0.9004\\\underline{29.17/0.8037}\\ 27.63/\underline{0.7385}\end{tabular}} &

{\begin{tabular}[c]{@{}c@{}}32.24/0.9007\\\underline{29.17}/0.8036\\\underline{27.67}/0.7383\end{tabular}}

& {\begin{tabular}[c]{@{}c@{}}\textbf{32.31/0.9011}\\ \textbf{29.26/0.8043} \\ \textbf{27.70/0.7399}\end{tabular}} \\ 
 \midrule
 {\textbf{Urban100}} & {\begin{tabular}[c]{@{}l@{}}$\times$2\\ $\times$3\\ $\times$4\end{tabular}} &  {\begin{tabular}[c]{@{}c@{}}30.76/0.9140\\27.14/0.8279\\25.18/0.7524\end{tabular}} & {\begin{tabular}[c]{@{}c@{}}31.23/0.9188\\27.53/0.8378\\25.44/0.7638\end{tabular}} & {\begin{tabular}[c]{@{}c@{}}31.31/0.9195\\27.56/0.8376\\25.50/0.7630\end{tabular}} & {\begin{tabular}[c]{@{}c@{}}31.92/0.9256\\28.06/0.8493\\26.07/0.7837\end{tabular}} & {\begin{tabular}[c]{@{}c@{}}31.41/0.9207\\26.41/0.8064\\24.60/0.7258\end{tabular}} & 
{\begin{tabular}[c]{@{}c@{}}32.17/0.9283\\28.17/0.8519\\26.04/0.7838\end{tabular}} &{\begin{tabular}[c]{@{}c@{}}32.01/0.9273\\28.11/0.8511\\26.11/0.7854\end{tabular}}  &
{\begin{tabular}[c]{@{}c@{}}32.38/0.9308\\28.26/0.8552\\26.16/0.7887\end{tabular}} &

{\begin{tabular}[c]{@{}c@{}}32.43/0.9302\\28.33/0.8538\\26.25/0.7873\end{tabular}} &

{\begin{tabular}[c]{@{}c@{}}\underline{32.52/0.9317}\\\underline{28.42/0.8578}\\ \underline{26.31}/0.7921\end{tabular}} &

{\begin{tabular}[c]{@{}c@{}}32.44/0.9311\\28.37/0.8572\\\underline{26.31/0.7923}\end{tabular}}

& {\begin{tabular}[c]{@{}c@{}}\textbf{32.60/0.9322}\\ \textbf{28.57/0.8581} \\ \textbf{26.39/0.7926}\end{tabular}} \\

\bottomrule

\end{tabular}
\end{adjustbox}

\label{tab:results}
\end{table*}
\begin{table*}[htb]
\caption{\small Average PSNR/SSIM for models with the same order of magnitude of parameters.
Scores shown for scale factor $\times3$ using \textbf{BD} degradation model. Best performance is \textbf{highlighted}, second best \underline{underlined}.}
\begin{adjustbox}{width=1\textwidth}

\setlength\arrayrulewidth{1.1pt}

\begin{tabular}{c@{~~}c@{~~~}c@{~~~}c@{~~~}c@{~~~}c@{~~}c@{~~}c@{~~}c@{~~}c@{~~~}c@{~~~}c}
\toprule
\textbf{DB}  & \begin{tabular}[c]{@{}c@{}} 
\textbf{Bicubic}\\  \end{tabular} & \begin{tabular}[c]{@{}c@{}} \textbf{SPMSR}\\  \end{tabular} & \begin{tabular}[c]{@{}c@{}} \textbf{SRCNN}\\  \end{tabular} & \begin{tabular}[c]{@{}c@{}} \textbf{FSRCNN}\\  \end{tabular} & \begin{tabular}[c]{@{}c@{}} \textbf{VDSR}\\  \end{tabular} & \begin{tabular}[c]{@{}c@{}} \textbf{IRCNN\_G}\\  \end{tabular} & \begin{tabular}[c]{@{}c@{}} \textbf{IRCNN\_C}\\  \end{tabular} & \begin{tabular}[c]{@{}c@{}} \textbf{SRMD(NF)}\\  \end{tabular} &
\begin{tabular}[c]{@{}c@{}} \textbf{MPRNet}\\ \end{tabular}&
\begin{tabular}[c]{@{}c@{}} \textbf{OverNet}\\ \end{tabular}&
\begin{tabular}[c]{@{}c@{}} \textbf{Ours}\\  \end{tabular} \\ \midrule 
\textbf{Set5} & \begin{tabular}[c]{@{}c@{}}28.34/0.8161\end{tabular} & \begin{tabular}[c]{@{}c@{}}32.21/0.9001 \end{tabular} & \begin{tabular}[c]{@{}c@{}}31.75/0.8988\end{tabular} & \begin{tabular}[c]{@{}c@{}}26.25/0.8130\end{tabular} & \begin{tabular}[c]{@{}c@{}}33.78/0.9198\end{tabular} & \begin{tabular}[c]{@{}c@{}}33.38/0.9182\end{tabular} & \begin{tabular}[c]{@{}c@{}}29.55/0.8246\end{tabular} & \begin{tabular}[c]{@{}c@{}}34.09/0.9242\end{tabular} &
\begin{tabular}[c]{@{}c@{}} 34.57/0.9278 \end{tabular}
&
\begin{tabular}[c]{@{}c@{}} \underline{34.59/0.9279}\end{tabular} & 
\begin{tabular}[c]{@{}c@{}} \textbf{34.66/0.9281}\end{tabular}\\ \midrule
\textbf{Set14}  & \begin{tabular}[c]{@{}c@{}}26.12/0.7106\end{tabular} & \begin{tabular}[c]{@{}c@{}}28.97/0.8205\end{tabular} & \begin{tabular}[c]{@{}c@{}}28.72/0.8024\end{tabular} & \begin{tabular}[c]{@{}c@{}}25.63/0.7312\end{tabular} & \begin{tabular}[c]{@{}c@{}}29.90/0.8369\end{tabular} & \begin{tabular}[c]{@{}c@{}}29.73/0.8292\end{tabular} & \begin{tabular}[c]{@{}c@{}}27.33/0.7135\end{tabular} & \begin{tabular}[c]{@{}c@{}}30.11/0.8304\end{tabular} &
\begin{tabular}[c]{@{}c@{}} \underline{30.47/0.8427} \end{tabular} &
\begin{tabular}[c]{@{}c@{}} 30.46/0.8310 \end{tabular} & 
\begin{tabular}[c]{@{}c@{}}\textbf{30.52}/\textbf{0.8429}\end{tabular} 
\\ \midrule
\textbf{B100} & \begin{tabular}[c]{@{}c@{}}26.02/0.6733\end{tabular} & \begin{tabular}[c]{@{}c@{}}28.13/0.7740 \end{tabular} & \begin{tabular}[c]{@{}c@{}}27.97/0.7921\end{tabular} & \begin{tabular}[c]{@{}c@{}}24.88/0.6850\end{tabular} & \begin{tabular}[c]{@{}c@{}}28.70/0.80035\end{tabular} & \begin{tabular}[c]{@{}c@{}}28.65/0.7922\end{tabular} & \begin{tabular}[c]{@{}c@{}}26.46/0.6572\end{tabular} & \begin{tabular}[c]{@{}c@{}}28.98/0.8009\end{tabular} &

\begin{tabular}[c]{@{}c@{}}\underline{29.19/0.8062}\end{tabular} &

\begin{tabular}[c]{@{}c@{}} 29.13/0.8060 \end{tabular} & \begin{tabular}[c]{@{}c@{}}\textbf{29.25}/\textbf{0.8065}\end{tabular} \\  \midrule
\textbf{Urban100} & \begin{tabular}[c]{@{}c@{}}23.20/0.6661\end{tabular} & \begin{tabular}[c]{@{}c@{}}25.84/0.7856 \end{tabular} & \begin{tabular}[c]{@{}c@{}}25.50/0.7812\end{tabular} & \begin{tabular}[c]{@{}c@{}}22.14/0.6815\end{tabular} & \begin{tabular}[c]{@{}c@{}}26.80/0.8191\end{tabular} & \begin{tabular}[c]{@{}c@{}}26.81/0.8189\end{tabular} & \begin{tabular}[c]{@{}c@{}}24.89/0.7172\end{tabular} & \begin{tabular}[c]{@{}c@{}}27.50/0.8370\end{tabular} &

\begin{tabular}[c]{@{}c@{}}\underline{28.31/0.8538}\end{tabular} &

\begin{tabular}[c]{@{}c@{}} 28.24/0.8485  \end{tabular} &  \begin{tabular}[c]{@{}c@{}}\textbf{28.39}/\textbf{0.8541}\\\end{tabular}\\  \midrule
\end{tabular}
\end{adjustbox}
\label{tab:Blurry}
\end{table*}

\subsection{Comparison with State-of-the-art Methods}

Simulating LR image with bicubic (BI) and blur-downscale (BD) degradation models is a common practice in SISR. In this section we report results using both strategies. We also present a comparative visual analysis of our proposed attention mechanism, and study \architecture{}'s computational performance with regard to state-of-the-art SISR models. 

\subsubsection{Results with BI Degradation Models}

We compare \architecture{} with 11 lightweight state-of-the-art methods \cite{kim2016accurate,tai2017image,tai2017memnet,ahn2018fast,li2019feedback,hui2019lightweight,zhao2020efficient, muqeet2020multi,luolatticenet,behjati2020overnet,mehri2020mprnet}. For fair comparison, we train models individually for each scale factor, including $\times$2, $\times$3 and $\times$4. We evaluate our model on different benchmarks using PSNR and SSIM as metrics. 

\begin{figure}[htb!]
\centering
\tiny
\begin{tabular}{ccccc}
\multirow{-6.715}{*}{\adjustbox{}{\includegraphics[width=.35\linewidth, height=2.99cm]{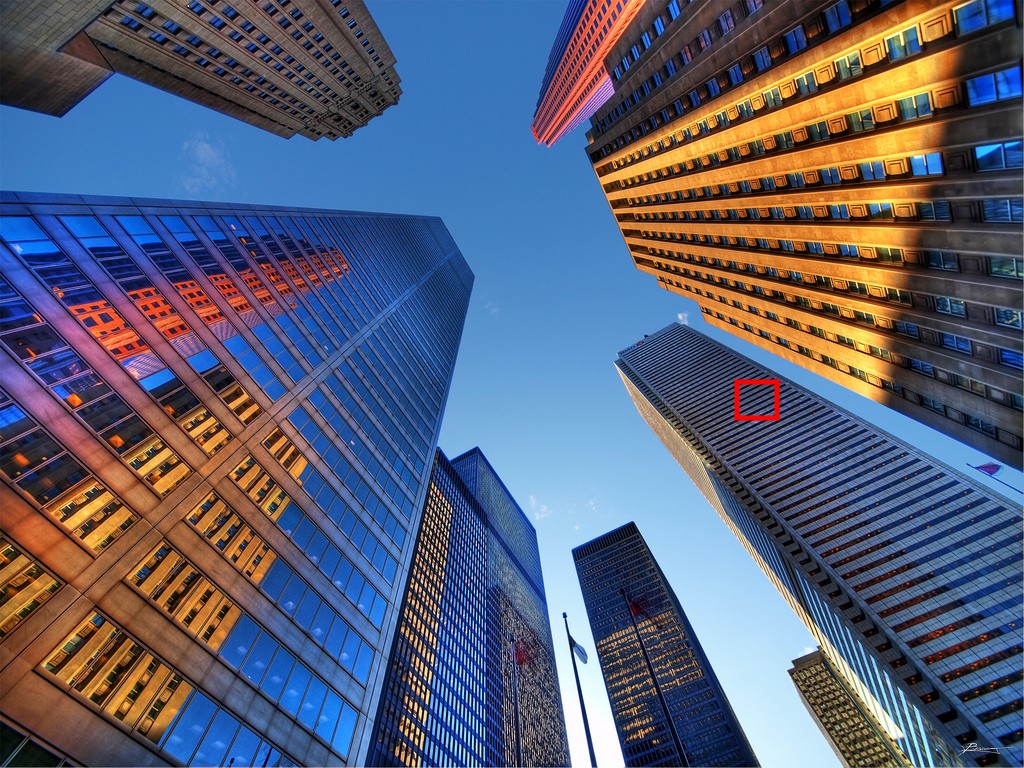}}} 
 & \includegraphics[width=.15\linewidth, height=1.360cm]{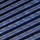} & \includegraphics[width=.15\linewidth, height=1.360cm]{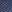} & \includegraphics[width=.15\linewidth, height=1.360cm]{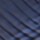} & \includegraphics[width=.15\linewidth, height=1.360cm]{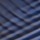} \\
 & HR & Bicubic & VDSR & MemNet \\
 & \includegraphics[width=.15\linewidth, height=1.360cm]{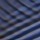} & \includegraphics[width=.15\linewidth, height=1.360cm]{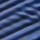} & \includegraphics[width=.15\linewidth, height=1.360cm]{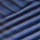} & \includegraphics[width=.15\linewidth, height=1.360cm]{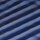} \\
 Img\_012 Urban100 & CARN & MPRNet & OverNet &  Ours \\
\end{tabular}

\begin{tabular}{ccccc}
\multirow{-6.715}{*}{\adjustbox{}{\includegraphics[width=.35\linewidth, height=2.99cm]{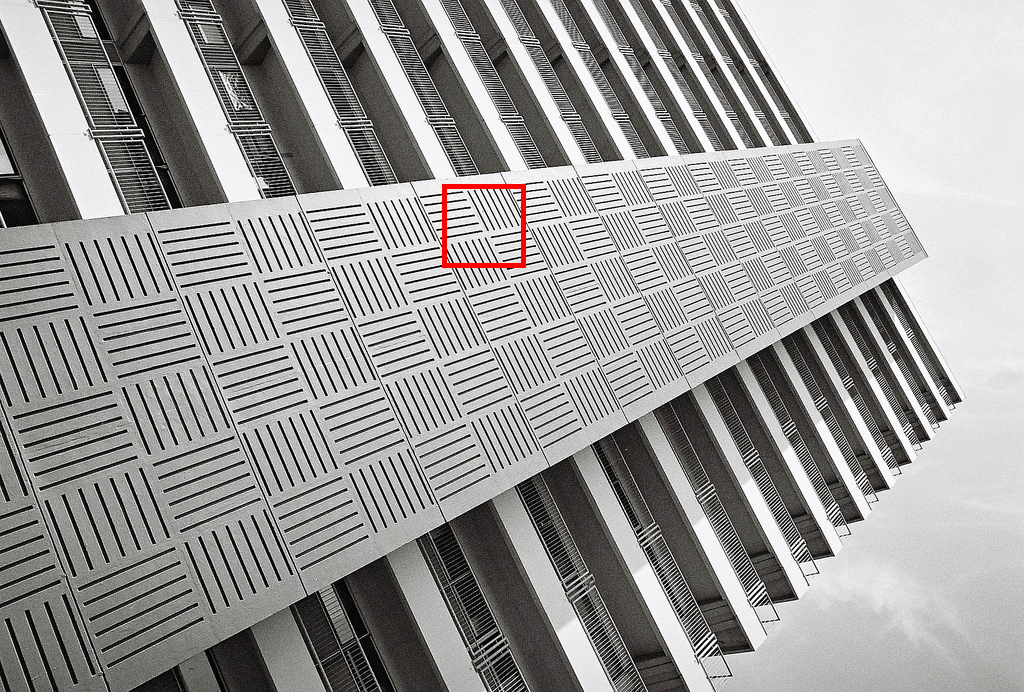}}} 
 & \includegraphics[width=.15\linewidth, height=1.360cm]{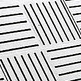} & \includegraphics[width=.15\linewidth, height=1.360cm]{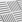} & \includegraphics[width=.15\linewidth, height=1.360cm]{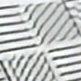} & \includegraphics[width=.15\linewidth, height=1.360cm]{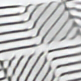} \\
 & HR & Bicubic & VDSR & MemNet \\
 & \includegraphics[width=.15\linewidth, height=1.360cm]{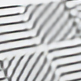} & \includegraphics[width=.15\linewidth, height=1.360cm]{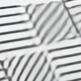} & \includegraphics[width=.15\linewidth, height=1.360cm]{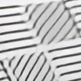} & \includegraphics[width=.15\linewidth, height=1.360cm]{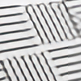} \\
 Img\_092 Urban100 & CARN & MPRNet & OverNet &  Ours \\
 \\
\end{tabular}
\caption{Visual results of \textbf{BI} degradation model for $\times$4 scale factor.}
\label{tab:bicubic}
\end{figure}

Table~\ref{tab:results} shows quantitative evaluation results, including the number of parameters for a more informative comparison. Note that in this table we only compare models that have a roughly similar number of parameters as ours. Compared to existing methods, and without using self-ensemble, our network is the best performing one at all scales. \architecture{} attains remarkable performance in all datasets including Urban100, which contains rich structure contents. This confirms that \architecture~is able to gradually aggregate, select and preserve relevant details across the whole network. These quantitative results also translate into more accurate reconstructions in Figure~\ref{tab:bicubic}, demonstrating that our method is able to generate clearer structures and details.

\subsubsection{Results with BD Degradation Model}
\begin{figure}[htb!]
\centering
\tiny
\begin{tabular}{ccccc}
\multirow{-6.715}{*}{\adjustbox{}{\includegraphics[width=.35\linewidth, height=2.99cm]{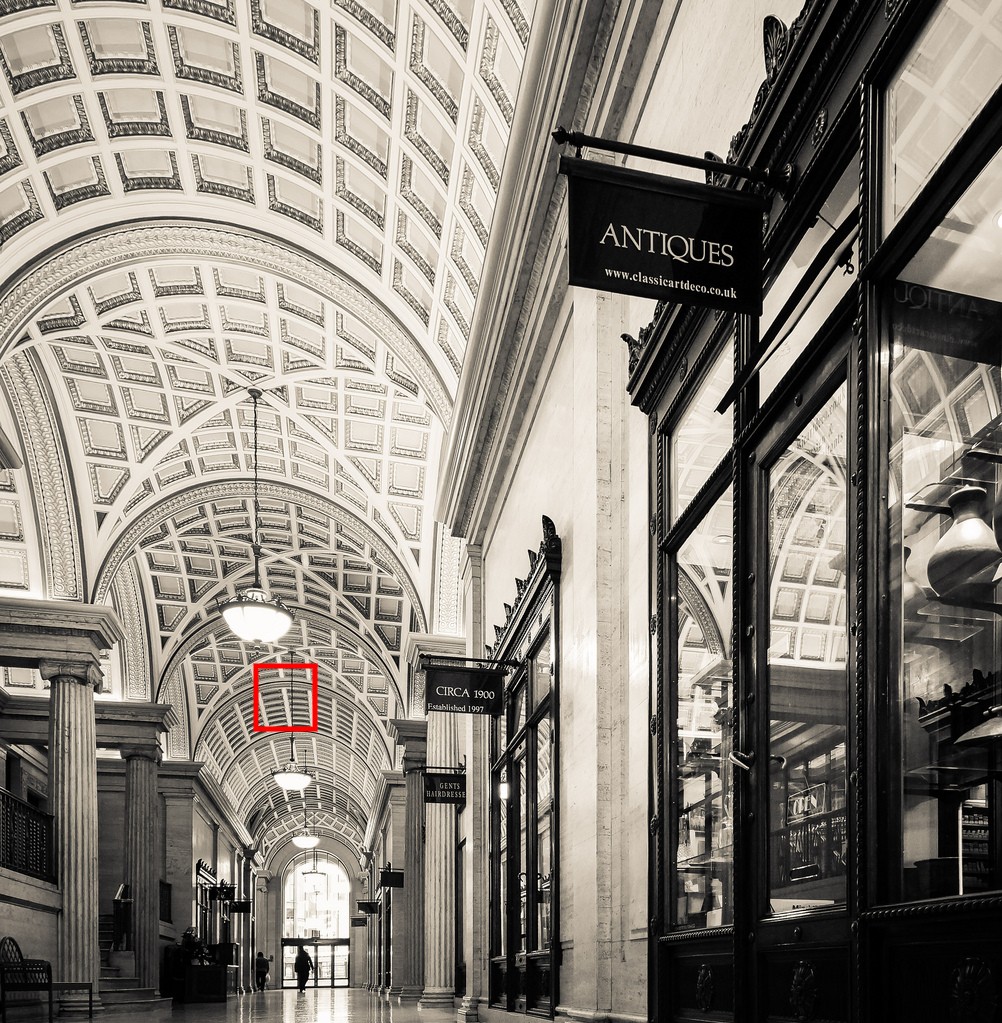}}} 
 & \includegraphics[width=.15\linewidth, height=1.360cm]{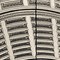} & \includegraphics[width=.15\linewidth, height=1.360cm]{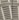} & \includegraphics[width=.15\linewidth, height=1.360cm]{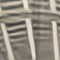} & \includegraphics[width=.15\linewidth, height=1.360cm]{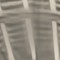} \\
 & HR & Bicubic & SRCNN &  FSCRNN\\
 & \includegraphics[width=.15\linewidth, height=1.360cm]{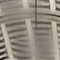} & \includegraphics[width=.15\linewidth, height=1.360cm]{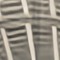} & \includegraphics[width=.15\linewidth, height=1.360cm]{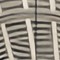} & \includegraphics[width=.15\linewidth, height=1.360cm]{Visualization/figures/pics_blurry/hr_croped.jpg} \\
 Img\_083 Urban100 & VDSR & MPRNet & OverNet &  Ours \\
 \\
\end{tabular}

\caption{Visual results of \textbf{BD} degradation model for $\times$3 scale factor.}
\label{tab:blurry_vis}
\end{figure}

Following \cite{zhang2018residual,dai2019second,behjati2020overnet,mehri2020mprnet}, we also provide the results after applying a BD degradation model. We compare our proposed method with 8 state-of-the-art-methods \cite{peleg2014statistical,dong2014learning,dong2016accelerating,kim2016accurate,zhang2017learning,tong2017image,behjati2020overnet,mehri2020mprnet}. Because of the mismatch of the degradation setups, SRCNN, FSRCNN and VDSR have been re-trained. As shown in Table \ref{tab:Blurry}, \architecture~achieves the best PSNR and SSIM scores compared to other SR methods on all the datasets. The consistently better results of \architecture{} indicate that our method can adapt well to scenarios with multiple degradation models. Figure~\ref{tab:blurry_vis} provides some visual results, where it becomes apparent that the quality of the reconstructions is superior for the proposed method. We found that \architecture{} is able to recover structured details that were missing the LR image by efficiently exploiting the feature hierarchy. 

\subsubsection{Visualization of Attention Mechanisms}
\begin{figure}[!htb]
\centering

\begin{tabular}{ccccc}
\footnotesize RB 1 &\footnotesize  RB 2 & \footnotesize RB 3 &\footnotesize  Output& \\ \includegraphics[width=.23\linewidth, height=2cm]{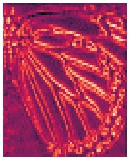}&
\includegraphics[width=.23\linewidth, height=2cm]{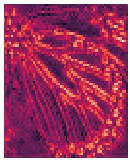} & \includegraphics[width=.23\linewidth, height=2cm]{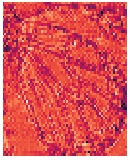} & \includegraphics[width=.23\linewidth, height=2cm]{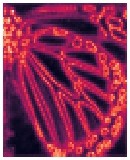}&
\includegraphics[width=.2\linewidth, height=2.1cm]{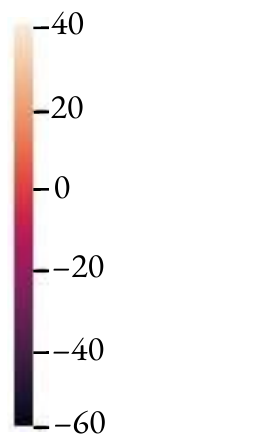}
\\
\end{tabular}

\begin{tabular}{ccccc}
\includegraphics[width=.23\linewidth, height=2cm]{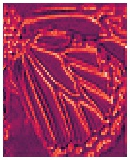}&
\includegraphics[width=.23\linewidth, height=2cm]{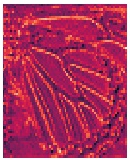} & \includegraphics[width=.23\linewidth, height=2cm]{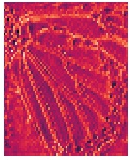} & \includegraphics[width=.23\linewidth, height=2cm]{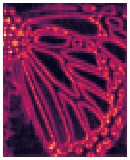}&
\includegraphics[width=.2\linewidth, height=2cm]{Figures/cbar.pdf}
\\
\end{tabular}

\caption{Average feature maps of residual blocks (RBs). \textbf{Top}: Attention is applied within the residual (classic approach). \textbf{Bottom}: Attention is applied outside the residual (our approach).}
\label{tab:feature_map}
\end{figure}

Figure~\ref{tab:feature_map} shows average feature maps of residual blocks, when attention mechanisms are applied within (in-place, top row) or outside (as in our RAFG, bottom row). 
This visualization shows 
how RAFGs 
are able to learn different representations, in contrast to in-place attention. In essence, each RAFG directs 
computations towards edges and details, thus obtaining a sharper representation at the output. In contrast, when using in-place attention, feature maps vary significantly from the first residual block to the last. As a result, edges and contours are outlined at the first layer, and smooth areas within the original image become suppressed at subsequent blocks. 
\setlength\tabcolsep{0.5pt}
\setlength\tabcolsep{0.5pt}

\subsubsection{Model Complexity and Running Time Analysis}
 
\begin{figure}[htb!]
    \centering
    \includegraphics[width=\linewidth, height=5.5cm]{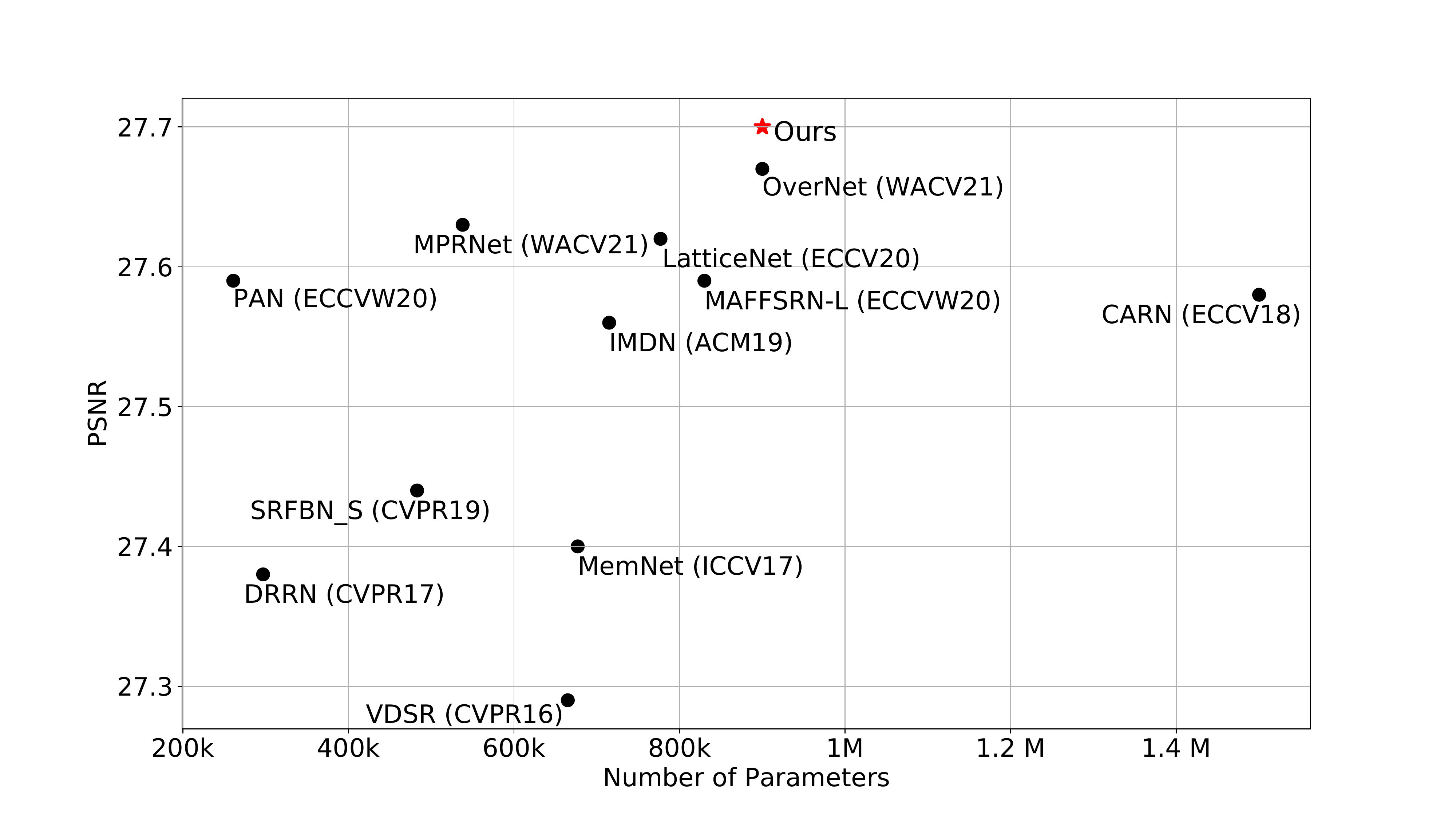}
    \caption{Comparing capacity vs performance for lightweight state-of-the-art SISR models B100 ($\times$4). The red star represents our method.}
    \label{fig:lightweight}
    \includegraphics[width=\linewidth, height=5.5cm]{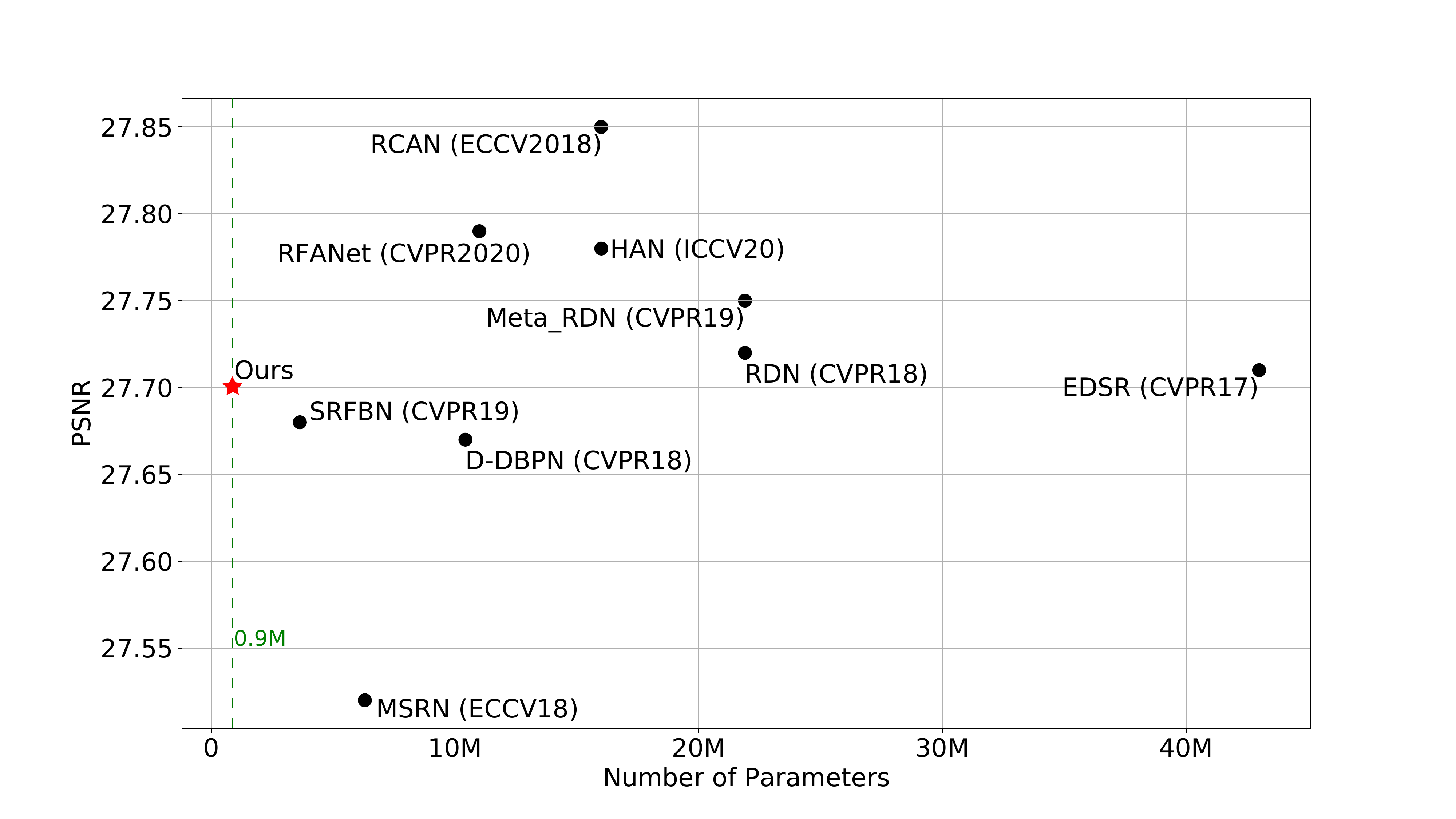}
    \caption{Comparing capacity vs performance for heavy state-of-the-art SISR models on B100 ($\times$4). The red star represents our method.}
    \label{fig:heavy}
  
\end{figure}

In Figure~\ref{fig:lightweight} and Figure~\ref{fig:heavy}, we compare \architecture{} against various lightweight and heavy benchmark algorithms respectively, in terms of network parameters and reconstruction PSNR, using the B100 dataset with a scale of $\times$4. It can be observed that \architecture{} achieves the best SR results among all lightweight networks. This demonstrates that our method achieves a good trade-off between number of parameters and reconstruction performance. Meanwhile, compared to the networks with larger number of parameters than \architecture{}, such as EDSR~\cite{lim2017enhanced} and RDN~\cite{zhang2018residual}, our proposed method achieves competitive or better results with only 2\% and 4\% the number of parameters of EDSR and RDN, respectively. 

Moreover, in Table~\ref{tab:eff} we compare the running time of \architecture{}  on Urban100  with five other lightweight state-of-the-art networks \cite{tai2017memnet,hui2019lightweight,behjati2020overnet,li2019feedback,mehri2020mprnet} using a scale factor of $\times$4. The running time of each network is evaluated  on the same machine with a NVIDIA 1080 Ti GPU. We observe that \architecture{} is the fastest, thus proving its efficiency.
\begin{table}[!htb]
\footnotesize
\caption{\small Average running time comparison on Urban100 for $\times 4$.}
\centering
\begin{tabular}{l@{~~~~~}c@{~~~~~~~}c@{~~~~~~~}c@{~~~~~~~}}
\toprule
\multirow{1}{*}[-.3em]{\begin{tabular}{c}\textbf{Model}\end{tabular}} & \multirow{1}{*}[-.3em]{\textbf{Parameters}}& \multirow{1}{*}[-.3em]{\textbf{Running Time (s)}} & \multirow{1}{*}[-.3em]{\textbf{PSNR}} \\  
\midrule
MemNet & 0.6M & 0.481 & 25.54  \\
SRFBN\_S & \textbf{0.4M} & 0.006 & 25.71 \\
IMDN  & 0.7M & 0.006 & 26.04  \\
MPRNet & 0.5M & 0.009 & 26.31 \\
OverNet & 0.9M & 0.004 & 26.31\\

\textbf{Ours} & 0.9M &\textbf{0.003} & \textbf{26.39}\\ \bottomrule
\end{tabular}%

\label{tab:eff}
\end{table}


\subsection{Ablation Study}

\subsubsection{Comparing Attention Schemes}
To demonstrate the effectiveness of our channel attention, we use HRAN as the basic network, and then replace our attention scheme with channel attention (CA)~\cite{wang2020eca}, efficient channel attention (ECA)~\cite{zhang2018image} and pixel-wise channel attention (PA)~\cite{zhao2020efficient}, respectively. Note that we only compare LCA with attention mechanisms that have a similar model complexity. As shown in Figure~\ref{tab:blocks}~(c), the channel attention module feeds from the residual block but splits out from it through a dedicated computational path. We train all the aforementioned methods using the same architectural pattern.

Table~\ref{CAs} compares the performance of all methods in terms of PSNR. It can be observed that LCA outperforms the rest of attention methods. Compared with CA, our proposed attention mechanism attains better performance by up to 0.9~dB, arguably because LCA avoids dimensionality reduction, thus preventing information loss~\cite{wang2020eca}.

\subsubsection{Effect of LCA on Existing Models}
This section investigates the impact of our proposed channel attention when applied to existing state-of-the-art models. Table~\ref{tab:sota_attention} shows the performance of applying attention mechanisms including CA~\cite{zhang2018image}, ECA~\cite{wang2020eca} and our proposed LCA on EDSR \cite{lim2017enhanced} and RCAN \cite{zhang2018image}. For fair comparison, all the methods are trained on their default settings, and attention mechanisms are performed \textit{in-place}, either at the end of the EDSR blocks or replacing RCAN's channel attention. We observe that all three attention mechanisms improve the performance of EDSR and RCAN, confirming the importance of attention in SISR tasks. 

We also observe that the proposed LCA outperforms CA and ECA on both architectures. We attribute the difference in performance with respect to CA to the fact that CA includes a two 1$\times$1 convolution bottleneck that may lose information during the dimensionality reduction process. ECA follows a similar approach, reducing dimensionality via a 1d convolution on the channel dimension. In contrast, our proposed LCA consists of a single 1$\times$1 convolution that directly maps the input into the attention weights. 

\subsubsection{Effect of the RAFG}
This section discusses the effect of each of the two dedicated computational paths in the proposed RAFG: feature banks and attention banks. \\
\noindent\textbf{Feature banks}. The proposed banks of residuals ensure that informative features from residual blocks can be hierarchically propagated without loss or interference. To verify its performance, we compared the use of hierarchical banks with ResNet, to evaluate the use of a regular architecture composed of several stacked residual blocks. Table~\ref{tab:FG} shows the results of the experiments, conducted on the Urban100 dataset with scale $\times$4.

The PSNR of ResNet is 26.03~dB on Urban100, a strong baseline for lightweight SISR methods. When deploying our hierarchical banks of residuals, the PSNR increases to 26.24~dB. Compared to ResNet, the proposed network only incorporates a 1$\times$1 convolution every three residual blocks, yet boosting the PSNR by 0.21~dB. We attribute this considerable improvement to the effectiveness of the proposed connectivity pattern, where the features in each residual block can be better utilized by the network. \\
\noindent\textbf{Lightweight Channel Attention}. Previous SISR approaches perform channel-attention \textit{in-place} within the residual blocks, whereas in this work we take the attention out and compute it in parallel. To verify the effectiveness of this approach, we compare the commonly used in-place attention with our proposed attention mechanism in Table~\ref{tab:in_out}. The two models in the table are identical, except that the top row one applies conventional attention within the residual as in Figure~\ref{tab:blocks}~(b), and the bottom row one implements our proposal as shown in Figure~\ref{tab:blocks}~(c). As it can be observed, our method achieves better results with only a few more parameters, due to the aggregation operation inside the attention feature bank.

\subsubsection{Effect of the Feature and Attention Banks}
We introduced feature and attention banks at the output of the network in order to grant the reconstruction module access to the full feature hierarchy. In this subsection we study the effectiveness of these banks by directly removing them and only keeping the output of the last RAFG and the long-range skip connection from the input. In Table~\ref{tab:banks} it can be observed that the proposed banks have a positive impact of almost 0.06~dB on the reconstruction performance of the network. This indicates that the banks successfully discard irrelevant information and distill valuable information from the entire feature hierarchy.

\setlength\tabcolsep{0.7pt}
\begin{table}[htb!]
\caption{\small Average PSNR for attention mechanisms with the same model complexity.}
\centering
\footnotesize
\begin{tabular}{c@{~~~~}l@{~~~~~~~}c@{~~~~~~~~}c@{~~~~~~~~}c@{~~~~~~}}
\hline
\textbf{Scale} &
  \textbf{Attentions} &
  \textbf{Params} &

  \begin{tabular}[c]{@{}c@{}}\textbf{Set14}\\ \textbf{PSNR/SSIM}\end{tabular} &

  \begin{tabular}[c]{@{}c@{}}\textbf{Urban100}\\ \textbf{PSNR/SSIM}\end{tabular} \\ \hline
\multirow{4}{*}{\textbf{$\times$4}} & \textbf{CA} & 864K & 28.69 & 26.30\\ 
                    & \textbf{ECA}  & 871K & 28.72 &  26.34 \\  
                    & \textbf{PA}  & 918K & 28.71 &  26.32 \\ 
                    & \textbf{LCA} & 918K  & \textbf{28.76} & \textbf{26.39}  \\ 
\hline
\end{tabular}
\label{CAs}
\end{table}
\begin{table}[!htb]
\centering
\footnotesize
\caption{\small Effect of attention modules on state-of-the-art methods with scale factor $\times$4.}
\label{tab:sota_attention}
\begin{tabular}{l@{~~}l@{~~~}c@{~~}c@{~~}c@{~~}c@{~~}c@{~~}c@{~~}c@{~~}}
\cline{1-9}
\multirow{3}{*}{\textbf{Name}} & \multicolumn{4}{c}{\textbf{EDSR}} & \multicolumn{4}{c}{\textbf{RCAN}}\\ 
\cmidrule(l){2-5} \cmidrule(l){6-9}
   & \textbf{Baseline} & \multicolumn{3}{c}{\textbf{Attention module}} & \textbf{Baseline} & \multicolumn{3}{c}{\textbf{Attention module}} \\ \cline{1-9} 
\textbf{CA} &  &   $\checkmark$    &     &     &     &       $\checkmark$     &         &        \\
\textbf{ECA}      &     &       &   $\checkmark$    &     &      &     &     $\checkmark$     &      \\
\textbf{LCA}   &  &    &    &  $\checkmark$   &       &        &      &   $\checkmark$      \\\hline
\textbf{Set5}   & 32.46   & 32.50    & 32.52    & \textbf{32.55}      & 32.63    & 32.67       & 32.69      & \textbf{32.73}      \\
\textbf{Urban100}    & 26.64    & 26.66      & 26.69  & \textbf{26.70}    & 26.82    & 26.86       & 26.87      & \textbf{26.91}     \\\hline
\end{tabular}
\end{table}
\setlength\tabcolsep{0.7pt}
\begin{table}[htb!]
\caption{\small Average PSNR for a regular ResNet architecture vs one using the proposed feature banks.}
\centering
\footnotesize
\begin{tabular}{c@{~~~~}l@{~~~~~~~}c@{~~~~~~~~}c@{~~~~~~~~}c@{~~~~~~~~}}
\hline
\textbf{Scale} &
  \textbf{methods} &
    \textbf{Params} &

  \begin{tabular}[c]{@{}c@{}}\textbf{Set14}\\ \textbf{PSNR/SSIM}\end{tabular} &
 \begin{tabular}[c]{@{}c@{}}\textbf{Urban100}\\ \textbf{PSNR/SSIM}\end{tabular} \\ \hline
\multirow{2}{*}{\textbf{$\times$4}} 
& \textbf{ResNet} & 749K & 28.52 & 26.03\\ 
& \textbf{Ours} & 814K & \textbf{28.76} &  \textbf{26.24} \\  

\hline
\end{tabular}
\label{tab:FG}
\end{table}
\setlength\tabcolsep{0.7pt}
\begin{table}[htb!]
\caption{\small Average PSNR with \textit{in-place} attention and our proposed attention.}
\centering
\footnotesize
\begin{tabular}{c@{~~~~}l@{~~~~~~~}c@{~~~~~~~~}c@{~~~~~~~~}c@{~~~~~~~~}}
\hline
\textbf{Scale} &
  \textbf{Methods} &
    \textbf{Params} &

  \begin{tabular}[c]{@{}c@{}}\textbf{Set14}\\ \textbf{PSNR/SSIM}\end{tabular} &

  \begin{tabular}[c]{@{}c@{}}\textbf{Urban100}\\ \textbf{PSNR/SSIM}\end{tabular} \\ \hline
\multirow{2}{*}{\textbf{$\times$4}} 
& \textbf{in-place} & 852K & 28.69 & 26.32\\ 
& \textbf{Ours} & 918K  & \textbf{28.76} &  \textbf{26.39} \\  

\hline
\end{tabular}
\label{tab:in_out}
\end{table}
\begin{table}[!htb]
\centering
\footnotesize
\caption{\small Average PSNR obtained with the HRAN model, using or not feature and attention banks at the end of the network.}
\label{tab:banks}
\begin{tabular}{c@{~~~}l@{~~~~~~~~}c@{~~~~~~~~~}c@{~~~~~~~~}}
\hline
\textbf{Scale}&\textbf{Datasets}   &   \textbf{HRAN w/o banks} & \textbf{HRAN w banks}\\\hline 
\multirow{3}{*}{\textbf{$\times$4}} 
&\textbf{Set14}    &    28.71        & \textbf{28.76}        \\
&\textbf{B100}     &    27.65          & \textbf{27.70}        \\
&\textbf{Urban100} &    26.34         & \textbf{26.39}       \\ \hline
\end{tabular}
\end{table}

\section{Conclusions}
We have introduced an efficient architecture (\architecture{}) for SISR. \architecture~is able to extract and preserve fine details from the whole feature hierarchy in order to reduce the degradation caused by successive residual aggregations. This is achieved through a novel Residual Attention Feature Group, which combines an efficient dense connectivity pattern with a lightweight channel attention path that is processed in parallel to the main residual computational path. We have empirically shown that the proposed architecture is effective at producing accurate high-resolution reconstructions of the input. As a result, \architecture~attains better reconstruction scores on all benchmarks than previous lightweight state-of-the-art models, while having a similar amount of parameters. Through a series of ablation experiments, we have demonstrated the effectiveness of the proposed attention mechanism as well as the RAFGs. We conclude that moving the attention operations outside of the residual blocks is beneficial to prevent the loss of information caused by \textit{in-place} attention. 
{\small
\bibliographystyle{abbrvnat}
\bibliography{egbib}
}

\end{document}